\documentclass[12pt,letterpaper]{article}

\usepackage[margin=1in]{geometry}
\usepackage{setspace}
\usepackage[english]{babel}
\usepackage[utf8]{inputenc}
\usepackage[T1]{fontenc}

\usepackage{amsmath}
\usepackage{amssymb}
\usepackage{bm}

\usepackage{mathptmx}
\usepackage{microtype}

\usepackage{graphicx}
\usepackage{booktabs}
\usepackage{tabularx}
\usepackage{float}

\usepackage{natbib}
\usepackage[colorlinks=true,linkcolor=black,citecolor=black,urlcolor=blue]{hyperref}

\usepackage{verbatim}
\usepackage{listings}
\begin{document}

\title{\Large \textbf{Social Contagion and Bank Runs: \\ An Agent-Based Model with LLM Depositors}}

\author{
Shreshth Rajan\thanks{Harvard College. Email: shreshthrajan@college.harvard.edu} \and
Christopher Rua\~{n}o\thanks{Harvard College. Email: cruano@college.harvard.edu}
}

\date{\today}

\maketitle

\begin{abstract}
\noindent Digital banking and online communication have made modern bank runs faster and more networked than the canonical queue-at-the-branch setting. While equilibrium models explain why strategic complementarities generate run risk, they offer limited guidance on how beliefs synchronize and propagate in real time. We develop a process-based agent-based model of bank runs that makes the information and coordination layer explicit. Banks follow cash-first withdrawal processing with discounted fire-sale liquidation, operational throughput limits, and an endogenous stress index. Depositors are heterogeneous in risk tolerance and in the weight placed on fundamentals versus social information, and they communicate on a heavy-tailed network tuned by Twitter/X activity during March 2023. Depositor behavior is generated by a constrained large language model (LLM) that maps each agent’s information set into a discrete withdraw or stay decision and an optional post that other agents can read; we validate this policy against laboratory coordination evidence and benchmark it to theoretical predictions. Across 4,900 parameter–seed configurations and full LLM simulations, three findings emerge. Increasing within-bank connectivity raises the likelihood and speed of withdrawal cascades holding fundamentals fixed. Cross-bank contagion exhibits a sharp phase transition in failure risk as information spillovers rise, with tipping behavior around spillover rates near 0.10. Depositor overlap and network amplification interact nonlinearly, so channels that are weak in isolation become powerful in combination. In an SVB–First Republic–regional scenario disciplined by crisis-era data, the model reproduces the observed ordering of failures and predicts substantially higher withdrawal rates among uninsured depositors. Overall, the results frame social correlation as a measurable amplifier of run risk alongside balance-sheet fundamentals, with implications for stress testing and supervisory monitoring.
\end{abstract}

\newpage

\section{Introduction}
\label{sec:introduction}

Economic models often perform best precisely where interaction and feedback are weakest. While the pursuit of closed-form solutions remains invaluable for clarity and elegance, it can be constraining when the phenomena of interest are driven by dense interactions among heterogeneous agents. In this respect, J. Doyne Farmer’s career has been influential in motivating alternative modeling approaches, most notably agent-based methods. Farmer helped build Prediction Company, an early quantitative trading firm grounded in the premise that even in systems as noisy and adversarial as financial markets, disciplined computational models can extract meaningful structure without claiming exact, closed-form truth. The relevant standard of success is not perfect description, but whether a model fed realistic inputs and operating on plausible heuristics generates dynamics that are decision-relevant, empirically falsifiable, and capable of improvement when they fail. Farmer’s later arguments for economics extend this logic: when interaction effects and nonlinear feedback loops are first-order, simulating a causal process with heterogeneous agents and institutionally grounded heuristics may be more informative than insisting on tractable equilibrium representations \citep{farmerfoley2009, axtell_2025}.

Bank-run theory sits uncomfortably with that aspiration. The foundational Diamond-Dybvig model explains why runs exist: demandable deposits provide liquidity insurance but create a coordination problem with multiple equilibria, including a run equilibrium \citep{diamond_dybvig_1983}. That insight is more than a curiosity, but rather the core logic of fragility under maturity transformation. But the same model also makes room for equilibrium selection driven by extrinsic belief shifts known as sunspots, in which the coordinating object has no intrinsic economic content beyond being commonly observed and commonly believed \citep{cass_shell_1983}. In principle, sunspots are coherent equilibrium devices. In practice, they are not reproducible as a causal explanation of modern runs since a sunspot is defined by the fact that it is unexpected and extrinsic, and equilibrium multiplicity does not tell the analyst when or how the economy will coordinate on a particular equilibrium ex ante. That gap matters for policy. Supervisors cannot stress-test a sunspot, measure it as an exposure, or target it with an intervention beyond blunt prescriptions that eliminate multiplicity altogether (e.g., full insurance or suspension both of which are outcomes that are meant to be used sparingly). As a result, equilibrium multiplicity often functions as an ex post label for crises rather than an operational tool for predicting which institutions are vulnerable and how fragility propagates.

The global-games literature is a serious response to this critique. By introducing small private signal noise, global games can deliver unique equilibrium selection and an ex ante run probability \citep{morris_shin_1998, goldstein_pauzner_2005}. This is an important corrective as with some adjustments equilibrium approaches can produce probabilistic predictions. The difficulty is that the probability result is only as credible as the assumed informational primitives—what signals agents receive, how noisy they are, what is common knowledge, and how these objects map to the real-world disclosure and communication environment. In modern episodes, information arrives through a mixture of formal announcements, private networks, and socially filtered communication that is itself shaped by actions during the run. Under these conditions, specifying and validating a single parametric signal structure that is stable across institutions and episodes is challenging, and small misspecifications can matter because selection hinges on informational details.

March 2023 illustrates why the information-and-coordination layer is much more than a sideshow. Using high-frequency payments data, \citet{cipriani_eisenbach_kovner_2024} trace deposit flows and identify 22 banks with run-like outflows depositors. Separately, \citet{cookson_2023} document that social media activity through Twitter/X played a catalytic role in the Silicon Valley Bank run and was associated with broader distress across regional banks during the same window. These papers do not imply that communication alone creates runs independent of fundamentals. They imply something more operational, that modern runs occur in an environment where correlation in beliefs and actions can be produced quickly by communication technologies and overlapping communities. That is exactly the type of mechanism that sunspot language gestures at but does not model in a way that can be measured, stress-tested, or used to design targeted mitigations.

This paper takes Farmer’s methodological bet seriously in a narrow, implementable form. We build a process-based agent-based model of bank runs that embeds a small set of explicit, interpretable heuristics to isolate this impact of social media and easy communication technology on bank runs. The bank’s balance-sheet block is governed by cash-first withdrawal processing with discounted fire-sale liquidation, a bounded per-round processing capacity, and an endogenous stress metric that responds to withdrawal intensity. The depositor block allows heterogeneity in risk tolerance and in trust placed on fundamentals versus social information. The communication block is a heterogeneous and heavy-tailed network within banks and (in the multi-bank extension) overlapping depositor bases across banks that we tune with real Twitter data. These choices are not derived as an optimal policy problem; they are meant to capture empirical regularities that matter for run dynamics in a way that is transparent, falsifiable, and extensible \citep{shleifer_vishny_1992, brunnermeier_pedersen_2009, axtell_2025}.

A key challenge is behavioral realism as once the model allows heterogeneous agents who respond to rich state descriptions and socially mediated narratives, hard-coded decision rules can become brittle and arbitrary. We therefore implement depositors as general social agents using a large language model (LLM) in the spirit of John J. Horton’s \emph{homo silicus}: an agent that maps a structured description of the economic environment into a discrete action and a short explanation \citep{horton_2023}. This differs from AI Economist approaches that train agents and planners via deep reinforcement learning within the environment \citep{zheng_2022}. We do not train a planner and we do not fine-tune the model on run data; instead, we constrain the agent through a strict action interface and evaluate it against laboratory evidence from a coordination experiment (and find that attempts at tuning actually decrease agent realism). The point is not that the LLM is a perfect cognitive model, but rather that it can serve as a disciplined, auditable behavioral prior in a simulator, provided it reproduces basic comparative statics and is subjected to ablations and robustness checks \citep{binz_2025}.

\subsection*{Contributions and preview}

The paper makes four contributions.

First, it advances a methodological claim: for modern bank runs, equilibrium multiplicity and sunspot selection are inadequate as a causal account of crisis dynamics and weak as a policy technology. A process model that specifies information sets, communication, and feedback loops provides a more operational framework for robustness assessment \citep{dogra_2024}.

Second, it formalizes social correlation as a distinct exposure channel. Overlap in depositor communities and their communication networks can synchronize withdrawals across banks, analogously to how correlated assets synchronize balance-sheet losses, but operating through beliefs and coordination rather than accounting identities \citep{cookson_2023}.

Third, it provides a behavioral validation exercise showing that an off-the-shelf LLM reproduces key qualitative patterns in a standard laboratory bank-run coordination setting. This supports using the LLM as a conservative baseline decision policy in the simulator \citep{arifovic_dejong_kopanyi_2024, garratt_keister_2009}.

Fourth, it delivers two headline simulation findings. In a single-bank environment, increasing depositor connectivity increases the likelihood and speed of run-like withdrawal cascades holding the stress path fixed. In a multi-bank environment designed around SVB, First Republic, and a regional control, cross-bank depositor overlap materially amplifies the propagation of distress from one institution to another, even when the second institution is mechanically healthier. The contribution is not a calibrated forecast; it is a proof of concept that a measurable social-structure object can be a first-order run amplifier in a process model that reflects real world phenomena \citep{cipriani_eisenbach_kovner_2024, cookson_2023}.

\subsection*{Roadmap}

Sections 2–3 position the paper and discipline the model: Section~2 situates our contribution in the literatures on bank runs, equilibrium selection, network contagion, and process-based modeling, while Section~3 summarizes the empirical anchors and theoretical benchmarks that guide our design. Section~4 then details the agent-based simulator, including bank balance-sheet mechanics, depositor heterogeneity, network formation, and the constrained LLM decision interface along with heuristic ablation baselines. Section~5 validates the LLM policy against laboratory coordination evidence. Section~6 reports the main simulation results in both single-bank and SVB/First Republic/multi-bank settings, emphasizing phase behavior, sensitivity surfaces, and channel decompositions. Section~7 concludes with implications for monitoring and stress testing, key limitations, and directions for extending the framework toward supervisory social-correlation risk metrics.

\section{Related literature and conceptual framing}
\label{sec:literature}

\subsection{Notable theoretical models of bank runs and their evolution}

The canonical starting point for any bank run paper is the Diamond-Dybvig model in which banks provide liquidity insurance by transforming illiquid assets into demandable deposits. Notably, though, it introduces the intuition of strategic complementarities: if depositors expect others to run, it becomes privately optimal to run, producing multiple equilibria \citep{diamond_dybvig_1983}. This framework is indispensable. It clarifies why maturity transformation is socially valuable but difficult to perform safely, and why runs can occur even when fundamentals are not obviously catastrophic.

At the same time, the equilibrium resolution of Diamond-Dybvig is where practical limitations begin. In the baseline model, equilibrium selection can be driven by extrinsic belief shifts, sunspots that select a run equilibrium at random, where the coordinating object has no intrinsic economic content beyond being commonly observed and commonly believed \citep{cass_shell_1983}. This is analytically elegant, but thin as a policy technology since by construction, a sunspot is not something supervisors can measure, stress, or mitigate in a targeted way. Preventing the bad equilibrium thus becomes a conceptual prescription rather than an operational one, unless the model provides a credible mapping from real-world information and communication into the belief shifts that coordinate runs.

The problem is therefore not that equilibrium logic is wrong, but that sunspot-style multiplicity is difficult to connect to depositor behavior at modern speed and says little about which objects that matter in practice. As a result, it can become hard to know which banks are vulnerable when runs ignite, how quickly they propagate, and what communication structures amplify them, even when we understand the game theory that might push an agent to pull their deposit as well. Multiplicity is a feature of the theory, but it does not automatically translate into a deployable framework for robustness assessment.

The global-games tradition responds directly to the equilibrium-selection problem by perturbing information with small noise. With noisy private signals and strategic complementarities, one can often obtain a unique equilibrium and derive an ex ante probability of a run \citep{morris_shin_1998, goldstein_pauzner_2005}. This is a major conceptual advance and directly relevant here, as it demonstrates that equilibrium approaches can, in principle, generate probabilistic predictions.

The cost is that the probability result is only as credible as the assumed signal environment, where there are choices around the distribution of noise, what is private versus public, what is common knowledge, and how agents aggregate signals into beliefs. In modern run settings, signals arrive through a mixture of official disclosures, informal private networks, and socially filtered communication that is itself shaped by the run. Even if one believes a global-games representation is correct in principle, specifying and validating the appropriate informational primitives in a March 2023-style episode is difficult, and small misspecifications can matter because selection hinges on knife-edge informational details. Global games therefore offer disciplined equilibrium selection, but the discipline comes from assumptions that are hard to measure and hard to stress-test when information is endogenous and network-mediated.

This paper does not reject global games. Rather, it emphasizes why selection matters as an important modeling object for real-world use, one that is our aim to study once we pivot to a process model that can incorporate richer information channels without committing to a single parametric signal structure upfront.

\subsection{Contagion mediators: why can connectivity be dangerous?}

A complementary literature to our models of bank runs studies fragility as a property of networks of financial exposures and correlated asset holdings. \citet{allen_gale_2000} demonstrate that when banks are linked through interbank claims, a liquidity shortfall at one institution can force asset sales or funding withdrawals at others, generating contagion. More recent work formalizes robust-yet-fragile network effects, in which increased connectivity initially improves stability by diversifying small, idiosyncratic shocks across many counterparties, but can ultimately make the system more vulnerable to large cascades once nonlinear constraints—such as balance-sheet limits, fire-sale externalities, or funding freezes—bind. In such networks, dense connections dampen minor disturbances yet provide efficient channels for the rapid propagation of distress when shocks are sufficiently large.\citep{acemoglu_ozdaglar_tahbazsalehi_2015, elliott_golub_jackson_2014, gai_kapadia_2010}.

These models are close cousins to the balance-sheet mechanisms in our simulator. When withdrawals force asset sales at discounts, mark-to-market losses can spill over to other institutions with correlated exposures. In the March 2023 episode, this channel was especially salient because banks were simultaneously exposed to valuation losses from the rapid rise in interest rates during 2022–23, which compressed the market value of long-duration securities across balance sheets and contributed to system-wide stress \citep{bis_2023}. The key distinction for us is where the network sits. Much of the systemic-risk literature focuses on interbank liabilities and direct balance-sheet linkages. Our emphasis is on a different but empirically salient network: communication and depositor overlap, which can synchronize withdrawals even when direct interbank exposures are limited.

The March 2023 episode is particularly informative because it was both fast and broad. Using high-frequency payments data, \citep{cipriani_eisenbach_kovner_2024} identify twenty-two banks that suffered run-like deposit outflows and show that runs were driven by a small number of large depositors and associated with weak balance-sheet characteristics. In a world in which coordination is made easier by digital banking and information moves at lightning speed, lots of the frictions that might once have helped to slow runs are now of little import. Retail queueing is no longer a relevant signal in an age of social media. 

\citep{cookson_2023} provide complementary evidence on the social media information layer of our model, as they argue that social media activity fueled the SVB run and that exposure to online communication was associated with broader distress across regional banks. The claim is not that social media creates runs out of nothing, but that it can act as an amplifier and coordinator—precisely the type of mechanism that is difficult to represent as a sunspot in an empirically testable and policy-relevant way. This evidence motivates the paper’s social correlation object. If depositor bases overlap socially through shared communities, influencers, or communication channels, then distress narratives can jump from one bank to another even when fundamentals differ. The model treats social overlap as an explicit risk factor, analogous to correlated asset exposures but operating through beliefs and coordination rather than accounting identities.

\subsection{Process versus equilibrium as a rationale for agent-based modeling}

Once communication-driven correlation is taken seriously as a mechanism, the central modeling question becomes how beliefs and actions co-evolve along a causal pathway rather than being pinned down by equilibrium consistency alone. \citep{dogra_2024} articulates a critique that crystallizes the methodological motivation for simulation. Equilibrium conditions impose consistency between beliefs and actions without specifying the causal mechanism by which the system arrives there. As a result, equilibrium models can yield paradoxes in causal interpretation, such as "immaculate revelation", in which agents appear to learn information no one initially possessed simply by observing equilibrium behavior.

In bank runs, this critique is more than philosophical. If the empirical object of interest is how panic starts, spreads, and accelerates—especially through endogenous communication—then the causal pathway is the mechanism. Treating that mechanism as an equilibrium-selection device or as an unspecified sunspot risks discarding precisely the structure needed for robustness assessment and policy design. A process model instead forces the analyst to specify what agents observe, how information flows, how actions feed back into fundamentals, and how the feedback loop evolves over time.

To that end, agent-based modeling (ABM) provides a computational framework for studying systems with heterogeneous agents, interaction effects, and nonlinear feedback—all features that often make closed-form equilibrium analysis brittle in crisis settings. \citet{axtell_2025} frames ABM as a way to relax representative-agent and equilibrium-sufficiency assumptions when micro heterogeneity and interaction are first order. \citep{farmer_2025} similarly argues for quantitative ABMs as a complement or alternative when standard models struggle to incorporate behavioral realism and interaction-driven dynamics.

This paper adopts that quantitative ABM perspective. It does not claim that ABM is intrinsically superior, but that it is a more operational container for studying run dynamics once heterogeneous interpretation, network-mediated communication, and fast feedback loops are admitted.

\subsection{LLMs as general social agents: promise, evidence, and the burden of validation}

A remaining challenge is how to model behavior inside an ABM without hard-coding brittle rules, especially when belief formation can be so variable in environments like social media. \citet{horton_2023} proposes using LLMs as simulated economic agents, arguing that they can serve as flexible behavioral engines conditioned on rich state descriptions while remaining testable against data.

There is also emerging evidence that language-model-based systems can be trained to predict human behavior across experimental settings. \citep{binz_2025} introduce Centaur, a model fine-tuned on experimental data that predicts behavior in tasks expressible in natural language. This evidence is informative but also raises a caution. Centaur is explicitly trained on experimental outcomes, whereas the present paper uses an off-the-shelf model embedded in a structured prompt interface. That difference increases the burden of validation and motivates treating LLM behavior as a conservative prior rather than as a settled structural model.

The approach here also differs from AI Economist-style work that trains agents and planners through deep reinforcement learning \citep{zheng_2022}. We do not train a planner. Instead, we embed a general-purpose language agent with a strict action interface and a policy function. This choice reduces training complexity and facilitates counterfactual experimentation, while increasing the importance of disciplined validation and sensitivity analysis.

\subsection{How the strands come together}

Taken together, the literature yields a coherent motivation. Diamond--Dybvig explains why runs exist, but multiplicity pushes key dynamics into belief selection that is hard to operationalize for supervision. Global games offer disciplined selection and probabilities, but rely on informational primitives that are difficult to justify and stress-test when communication is endogenous. Network models show that connectivity can be stabilizing or destabilizing and that systemic outcomes can exhibit tipping behavior. The 2023 episode provides direct evidence that runs were fast, broad, and plausibly amplified by communication channels. Dogra’s critique provides a methodological rationale for modeling the causal pathway rather than relying on equilibrium consistency as a causal explanation.

This paper’s contribution is to use a process-based agent-based model that makes communication and overlapping depositor bases explicit, and to study how these features amplify run risk alongside fundamentals. The goal is not to discard equilibrium theory, but to move beyond sunspot-style selection as the end of the story and toward a framework that can ultimately support deployable robustness assessment for social correlation risk in digital-age bank runs.

\section{Data}
\label{sec:data}

Although the paper’s core contribution is a simulation model, its inputs are disciplined by three empirical anchors: (i) bank balance-sheet characteristics that shaped vulnerability in March 2023, (ii) evidence on communication and coordination, especially via social media, and (iii) experimental evidence used to validate the LLM decision policy and behavioral realism. This section documents what is measured or taken from external sources versus what is imposed as a scenario design choice.

\subsection{Bank fundamentals and scenario inputs}

\subsubsection{Balance-sheet inputs used in the simulator}
\label{subsubsec:sim_inputs}

Each bank $b$ is initialized with total deposits $D_{b0}$ and a split between liquid assets $L_{b0}$ and long-term illiquid assets $A_{b0}$. In the implementation, these objects are set via ratios such as liquidity ratios and long-term-asset ratios rather than attempting to reproduce full regulatory balance sheets. The model also specifies:

\begin{itemize}
    \item \textbf{Fire-sale discount} $h_b$ governing liquidation proceeds. When withdrawals exceed available liquidity, the bank liquidates long-term assets at an effective discount (haircut) and raises only $h_b$ dollars of cash per dollar of assets sold. This is a reduced-form way to represent funding-liquidity stress and forced sales at dislocated prices \citep{shleifer_vishny_1992, brunnermeier_pedersen_2009}. In the baseline multi-bank calibration, we set modest discounts in the 10--15\% range (SVB $h_b=0.85$, First Republic $h_b=0.88$, Regional $h_b=0.90$). These are conservative relative to regulatory liquidity frameworks and stress-testing conventions that explicitly emphasize large haircuts under severe stress (and apply substantial valuation haircuts for less liquid asset categories) such as Basel III, which assigns discounts up to 50\% for some assets. At the same time, empirical fire-sale discounts can vary widely by asset class and market conditions; documented episodes include dislocations on the order of 5-10\% even for relatively standard securities in forced-sale environments \citep{covalstafford2007, shleifervishny2011}. Our cross-bank ranking of $h_b$ is intentional: it reflects differences in effective liquidation severity driven by preparedness and liquidation horizon. In the SVB episode, supervisory reviews emphasize that the bank could not mobilize enough cash or collateral quickly and was inadequately prepared to access contingent funding sources, which is consistent with a lower effective $h_b$ under stress \citep{fed_svb_review2023}. More generally, allowing institutions more time to liquidate attenuates fire-sale losses, so $h_b$ can be interpreted as partially capturing (reduced-form) differences in liquidation horizon across scenarios \citep{cont_schaanning2017}.

    \item \textbf{Initial perceived default risk} $pd_{b0}$, together with the rule for PD updating (Section~\ref{sec:methodology}). We treat $pd_{bt}\in[0,1]$ as a compact, observable stress index that depositors condition on. Baseline values (SVB $0.05$, First Republic $0.03$, Regional $0.01$) are scenario initializations chosen to encode the intended ordering at $t=0$ while leaving room for endogenous feedback from withdrawals; importantly, these initial values sit below the elevated thresholds embedded in the multi-bank prompt (Appendix~A), so that subsequent dynamics reflect shocks, spillovers, and coordination rather than starting from an already severe state.

    \item \textbf{Uninsured-deposit share}, used to label agents as insured versus uninsured in prompts. Uninsured status is sampled at the depositor level using a bank-specific uninsured share, because uninsured exposure changes both incentives and coordination intensity in run environments. SVB is calibrated near the supervisory-reported year-end 2022 uninsured share (94\%), implemented as 0.93 in the baseline; First Republic is set to 0.68, matching FDIC OIG estimates from call report data at 12/31/2022 (68\%) \citep{fed_svb_review2023, fdic_oig_frc_mlr_2023}. The Regional control is set to 0.40 as a stylized ``more typical'' liability mix, consistent with the fact that SVB and Signature were outliers in their reliance on uninsured deposits relative to most banks \citep{fed_fsr_2023_funding_risks}.

    \item \textbf{FDIC-style seizure thresholds}, defined as cumulative withdrawal-fraction cutoffs that mark banks failed or seized even in the absence of a mechanical payout shortfall. This is a model closure rule rather than a literal FDIC trigger: it captures the idea that beyond some scale of outflows, a bank becomes non-viable (or is resolved) even if it can mechanically process some withdrawals via asset sales. The SVB threshold is set to 0.25 to match the order of magnitude of the one-day outflow documented in the Federal Reserve's postmortem (over \$40 billion on March 9, preceding closure on March 10) \citep{fed_svb_review2023}. First Republic's threshold is set higher (0.35) to reflect greater time and support in the scenario, including the real-world consortium deposit placement intended to stabilize the bank-before resolution becomes inevitable \citep{fdic_oig_frc_mlr_2023}. The Regional control is higher still (0.40), representing a bank that is comparatively insulated and slower to reach a resolution boundary.
\end{itemize}

The intent is to capture salient dimensions of fragility liquidity shortfall exposure, mark-to-market sensitivity through forced sales, and a high share of uninsured deposits without claiming that the simulator reproduces each institution’s full accounting identity. Table~\ref{tab:bank_params} summarizes the baseline multi-bank parameterization.

\begin{table}[t]
\centering
\caption{Baseline multi-bank parameters. FDIC-style seizure threshold is a model closure rule: a bank is flagged failed/seized once cumulative withdrawals exceed the stated fraction of initial deposits (in addition to mechanical payout failure).}
\label{tab:bank_params}
\begin{tabularx}{\textwidth}{lcccc}
\toprule
Bank & Fire-sale $h_b$ & Initial $pd_{b0}$ & Uninsured share & Seizure threshold \\
\midrule
SVB            & 0.85 & 0.05 & 0.93 & 0.25 \\
First Republic & 0.88 & 0.03 & 0.68 & 0.35 \\
Regional       & 0.90 & 0.01 & 0.40 & 0.40 \\
\bottomrule
\end{tabularx}
\end{table}

\subsubsection{External evidence motivating parameterization}
Two empirical facts from the 2023 episode are especially important for these modeling choices. First, SVB’s depositor base was overwhelmingly uninsured, implying unusually strong coordination incentives and loss exposure. Second, First Republic also exhibited exceptionally high uninsured-deposit exposure but still to a notably lesser extent than SVB itself. These facts enter the simulation in two places: (i) sampling which agents are insured versus uninsured, which affects incentives and prompt content, and (ii) motivating the scenario design in which a bank with meaningfully different fundamentals (First Republic versus SVB) can nonetheless experience acute distress under correlated withdrawals.

A broader empirical anchor for the systemic nature of the March 2023 episode is evidence from high-frequency payments data showing that twenty-two banks experienced run-like deposit outflows, driven by a small number of large depositors and associated with weak balance-sheet characteristics \citep{cipriani_eisenbach_kovner_2024}. This motivates modeling run risk as a continuum of withdrawal intensity and timing rather than as a purely binary failure event.

\subsubsection{Scripted timeline events}
In the multi-bank environment, parts of the SVB and First Republic stress paths are scripted (e.g., deterministic PD drift and a discrete jump for SVB; a liquidity injection and PD reduction for First Republic at a fixed round) to match the events of the Panic of 2023 in our multi-bank simulations. These interventions are scenario design choices, used to anchor a baseline stress narrative so that the analysis can focus on the incremental role of social correlation and overlap rather than attempting to endogenize the entire 2023 macro-financial environment. The First Republic support event is motivated by the real-world consortium placement of \$30 billion in uninsured deposits intended to stabilize the bank and restore market confidence \citep{fdic_oig_frc_mlr_2023}. More generally, the design follows a common practice in stress testing: fix a plausible macro/episode path to isolate the marginal contribution of a specific amplification mechanism---here, communication-driven correlation and overlapping depositor communities---to run dynamics \citep{cipriani_eisenbach_kovner_2024, cont_schaanning2017}.

\subsection{Communication and social-media evidence}

\begin{figure}[!htbp]
\centering
\includegraphics[width=0.85\textwidth]{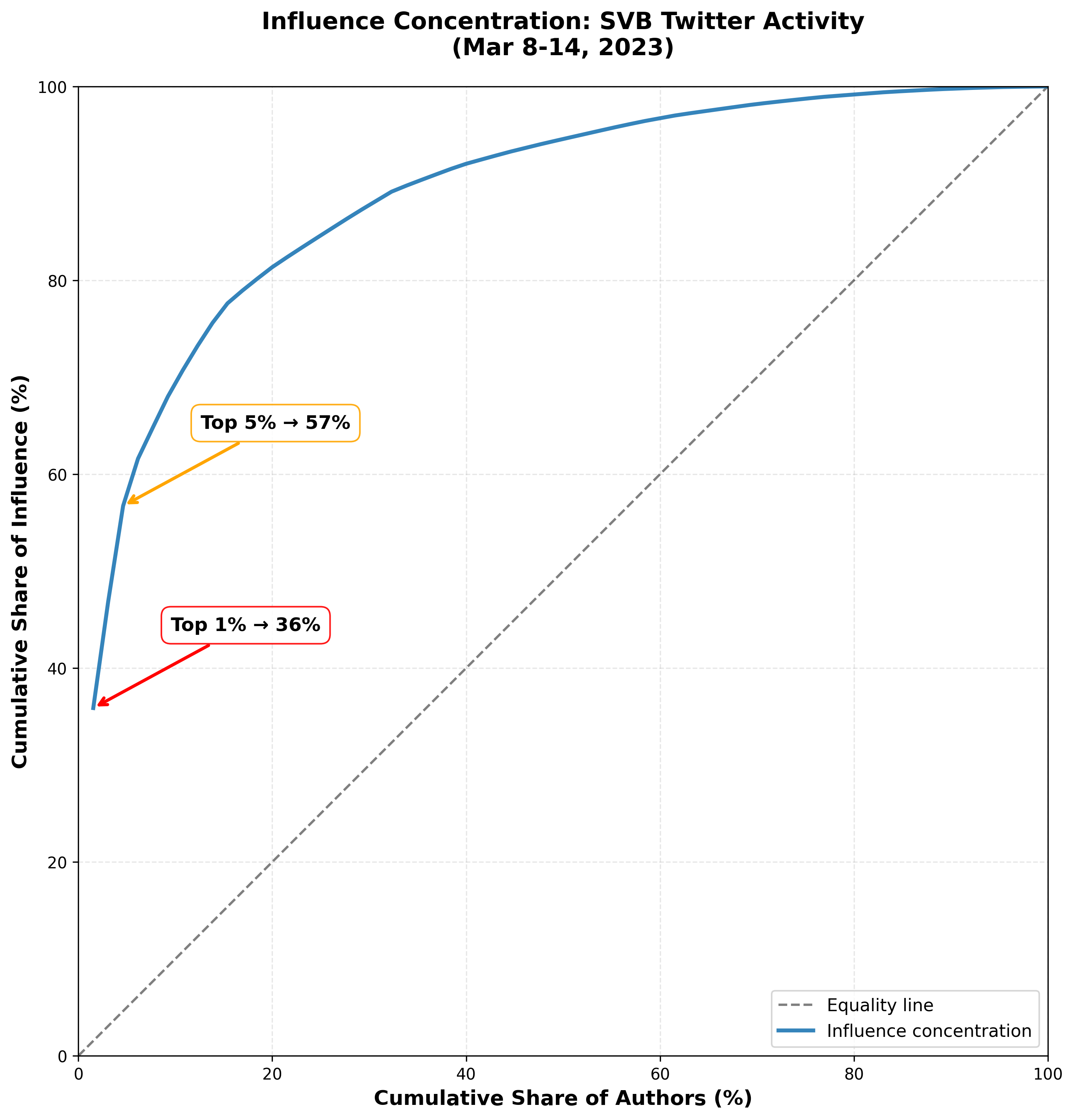}
\caption{Influence concentration in SVB-related retweet activity (Cookson et al.\ replication). A small fraction of accounts generates a disproportionate share of amplification, motivating a hub-dominated diffusion process and a core--periphery social graph within the SVB depositor network.}
\label{fig:influence_concentration}
\end{figure}

\begin{figure}[!htbp]
\centering
\includegraphics[width=0.95\textwidth]{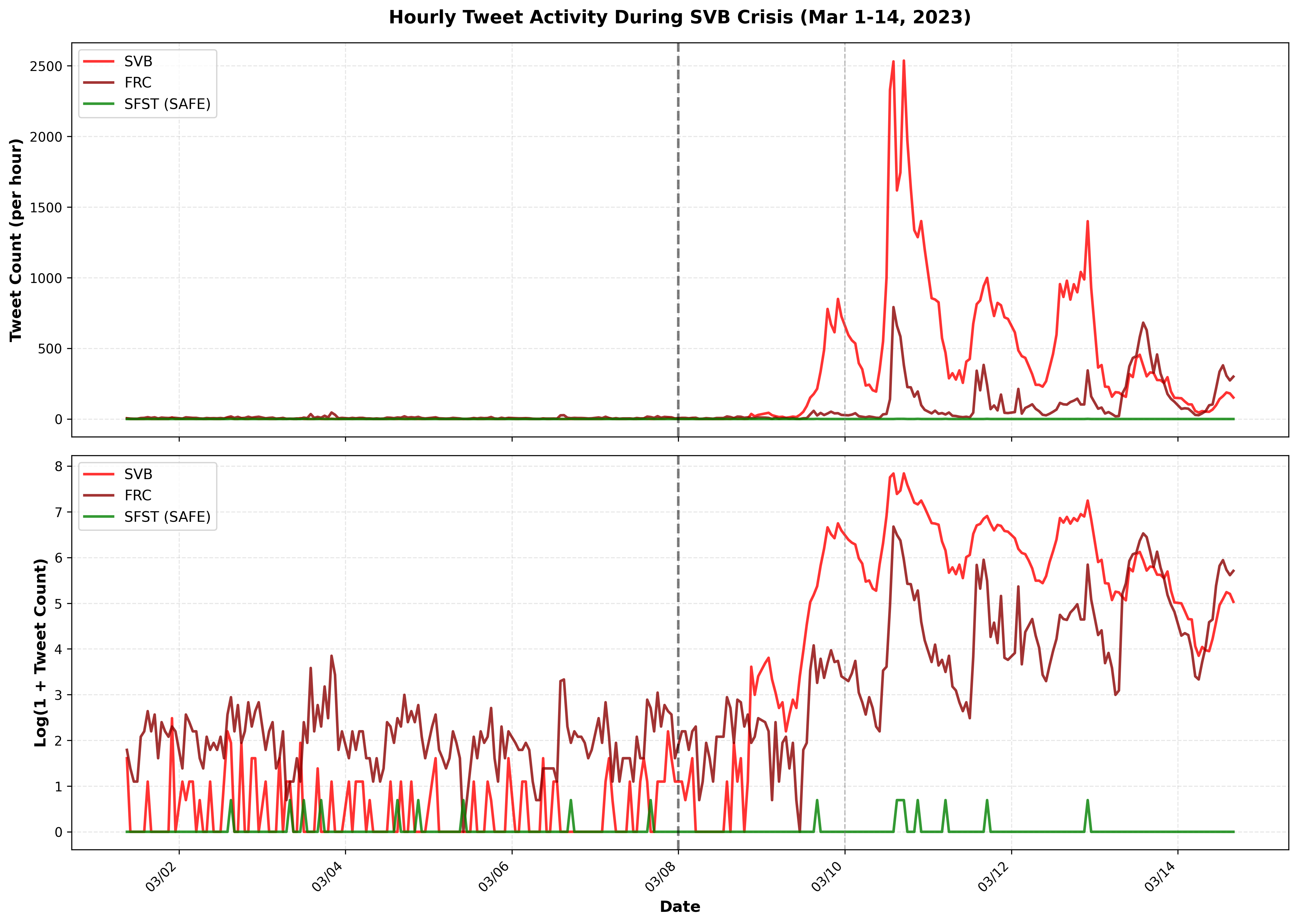}
\caption{Hourly Twitter/X activity during the crisis window: SVB-topic activity (aggregating \texttt{\$SIVB} and \texttt{\$SVB} where available) versus other tickers. FRC co-moves strongly with the SVB topic, while SFST remains comparatively insulated in both levels and log scale, motivating the use of $q_{\text{safe}}$ to discount SVB-origin spillovers to the safe-bank benchmark.}
\label{fig:tweet_timeseries}
\end{figure}

\begin{figure}[!htbp]
\centering
\includegraphics[width=0.90\textwidth]{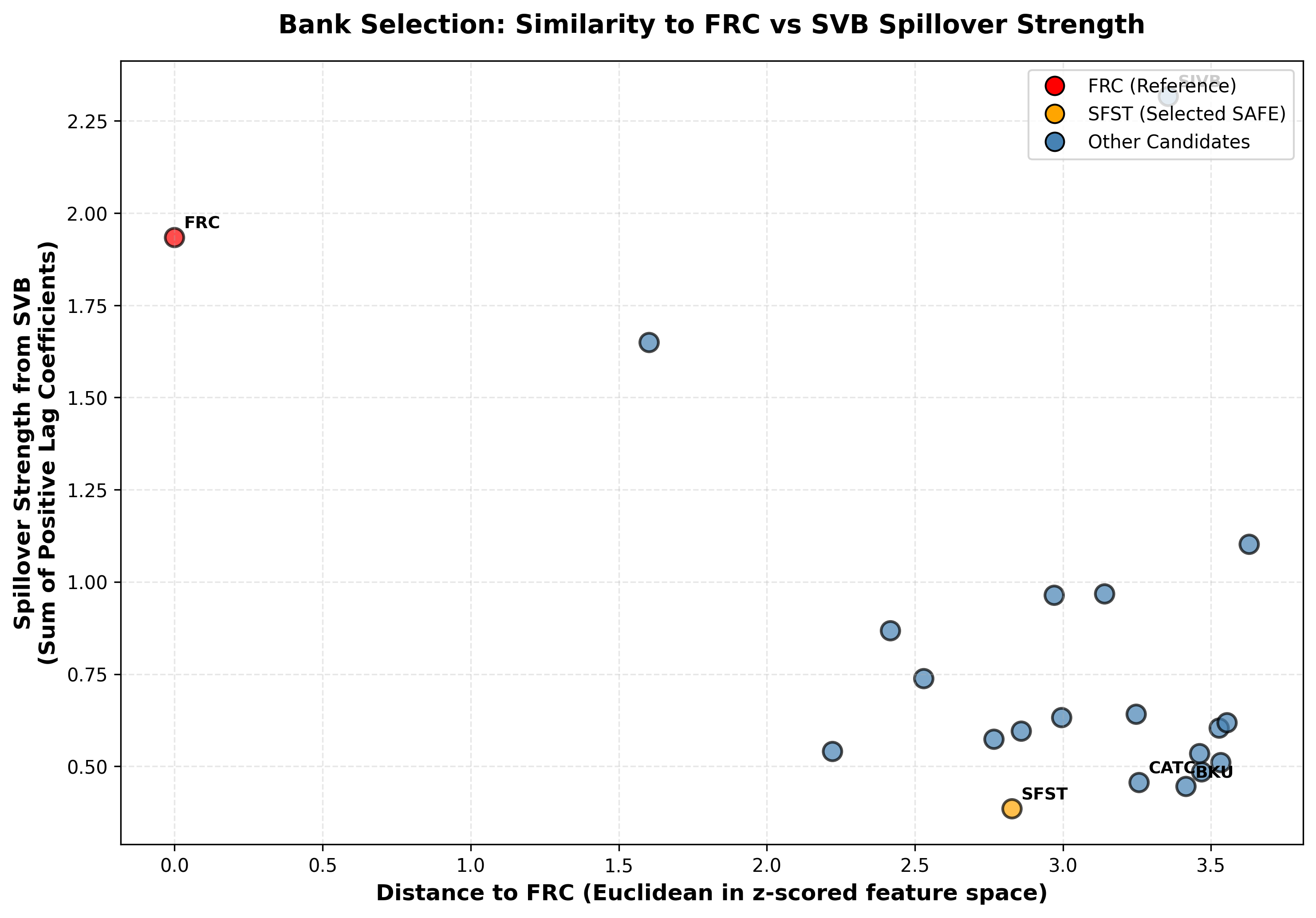}
\caption{Selecting the ``safe'' regional benchmark. Scatter of (fundamental distance to FRC) versus estimated SVB$\rightarrow$bank spillover strength (distributed-lag sum of positive coefficients). SFST is selected as the bank closest to FRC in fundamentals while exhibiting minimal SVB-topic spillover, and is used to define $q_{\text{safe}}$.}
\label{fig:distance_vs_spillover}
\end{figure}

We discipline the model’s communication layer using the Cookson et al.\ (2024) replication data that contains all of the tweets from the period surrounding the Panic of 2023 that mentions Silicon Valley Bank using either the stock ticker, SIVB, or the abbreviation SVB. Using this data, we can tune the parameters of our social network as we build out our social media modeling device, something that is further detailed in Section 4. To do so, we extract moments that map directly into within-bank network concentration and cross-bank spillover strength \citep{cookson_2023}. Figure~\ref{fig:influence_concentration} shows that SVB-related attention is highly concentrated: a small fraction of accounts generates a disproportionate share of retweet amplification, motivating a hub-dominated diffusion process and a core-periphery network structure within the SVB depositor graph. To identify cross-bank propagation, we use the hourly bank panel and construct an SVB topic series (aggregating SIVB and, where available, SVB mentions), then compare the time path of SVB-topic activity to other tickers during the crisis window. Figure~\ref{fig:tweet_timeseries} illustrates that FRC co-moves strongly with the SVB topic in quick succession, while SFST remains largely insulated in both levels and log scale. We formalize this asymmetry in a Twitter tuning pipeline that runs distributed-lag regressions of hourly ticker activity on lagged SVB-topic activity and defines spillover strength as the sum of positive lag coefficients. We then select SFST as the representative safe regional bank by restricting to banks similar to FRC in balance-sheet characteristics and choosing the bank with minimal SVB spillover; Figure~\ref{fig:distance_vs_spillover} shows SFST’s position as somewhat close in fundamentals yet weakly connected in social-media spillovers. The resulting ratio of spillover strengths defines $q_{\text{safe}}$, the relative cross-bank coupling parameter:
\[
q_{\text{safe}} \equiv \frac{\text{spillover}(\text{SVB}\rightarrow \text{SAFE})}{\text{spillover}(\text{SVB}\rightarrow \text{FRC})},
\]
which is approximately 0.20 in our baseline calibration and is written out alongside bank-specific activity multipliers as drift moments for the ABM. In the simulation configuration, this estimate enters the multi-bank system through two channels: it scales the density and effective weight of SVB$\leftrightarrow$SAFE bridge connections (with calibrated bridge fractions set to 5\% for SVB$\leftrightarrow$FRC and 0.1\% for SVB$\leftrightarrow$SAFE, informed by the same overlap analysis), and it governs the contagion engine’s cross-bank transmission. Specifically, $q_{\text{safe}}$ discounts informational spillovers so the SAFE bank receives a smaller probability-of-distress shock from SVB-origin narratives, and it dampens crisis scrutiny multipliers in the fear/withdrawal equation, capturing a flight-to-safety pattern in which attention and panic propagate strongly to FRC but only weakly to the regional benchmark.

\subsection{LLM policy: inputs, documentation, and a conservative validation check}

In our model, depositor behavior is generated by an LLM that serves as a behavioral policy function: it maps a depositor’s information set into a discrete action (withdraw or stay) and, optionally, a short message to post to the social feed. Concretely, behavior depends on the model specification (gpt-4.1-mini), temperature (0.7), and token budget (512); a prompt template (Appendix~\ref{app:prompts}) that deterministically formats the depositor’s state (numeric fundamentals, insurance status, a qualitative summary, last-round withdrawal rate, and lagged neighbor posts); and the simulation seed, which fixes shocks, networks, and agent traits and therefore fixes the prompt inputs. Because the LLM is part of the data-generating process in this mode, these configuration choices are not implementation details: they define the behavioral mapping and are therefore recorded explicitly and stress-tested via ablations (LLM vs.\ heuristic baseline; feed on vs.\ off; overlap scaling).

We do not defend this choice on the claim that LLMs are accurate models of human cognition. The narrower claim is that, in coordination environments, an untuned LLM can provide a useful structural prior that captures the direction of key comparative statics while remaining easy to embed in a process-based simulator \citep{horton_2023, binz_2025}. To discipline this component, we conduct a deliberately conservative validation exercise using a single laboratory bank-run design with a clean mapping to our action space. We validate against a two-bank “pre-deposit” run experiment where subjects choose between a higher-return but fragile bank (“Round”) and a lower-return but safer bank (“Square”), and then decide whether to withdraw early or wait. Bank vulnerability varies across low/medium/high treatments \citep{garratt_keister_2009, arifovic_dejong_kopanyi_2024}. Focusing on first-round decisions only (to abstract from learning and path dependence), each choice corresponds to one of four actions:
\[
a \in \{\text{Round--Withdraw},\ \text{Round--Wait},\ \text{Square--Withdraw},\ \text{Square--Wait}\}.
\]
For each vulnerability regime we query the LLM repeatedly using a prompt that describes the payoff tradeoff and the coordination externality, but contains no empirical frequencies, subject-level data, or equilibrium labels. We aggregate sampled outputs into an empirical action distribution and compare the resulting comparative statics to the experimental patterns.

\begin{figure}[!htbp]
  \centering
  \includegraphics[width=\linewidth]{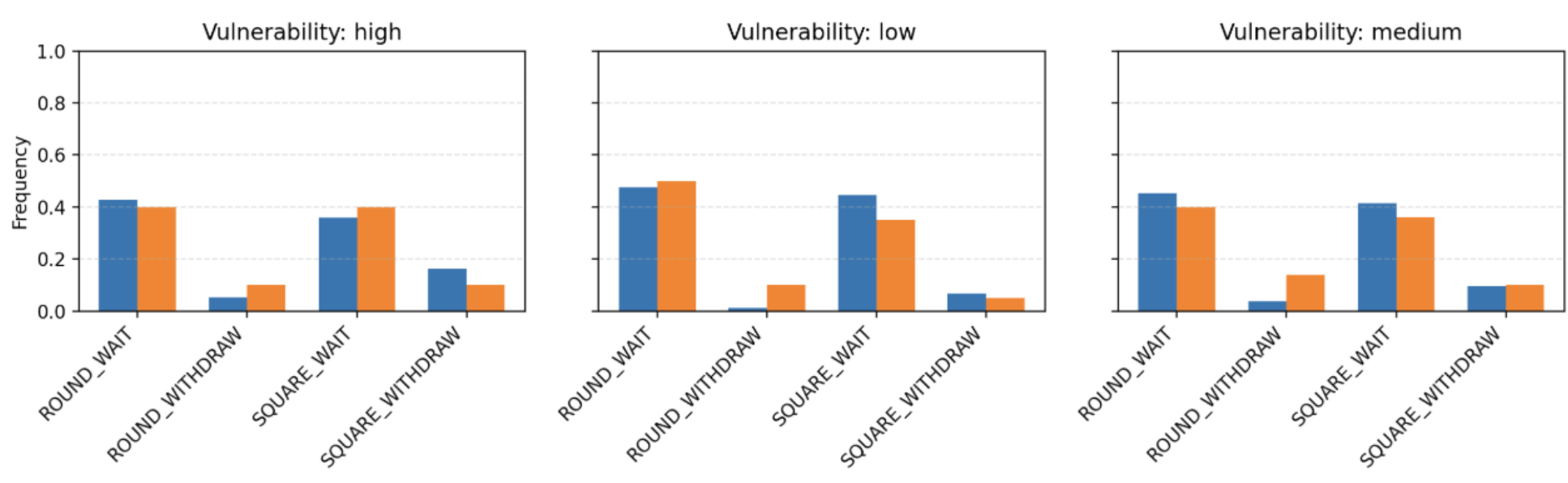}
  \caption{\textbf{LLM behavioral validation (first-round choices).} The figure compares the LLM’s sampled first-round action frequencies to the experimental benchmark across low/medium/high vulnerability regimes for the Round (high return, fragile) vs.\ Square (lower return, safer) banks. The LLM reproduces the qualitative comparative statics: as vulnerability rises, mass shifts toward the safer bank and early withdrawal becomes more prevalent primarily in the high-vulnerability regime.}
  \label{fig:llm_validation}
\end{figure}

Figure~\ref{fig:llm_validation} summarizes the result: in low and medium vulnerability regimes, the LLM places most mass on waiting, consistent with patience when coordination is plausible; as vulnerability increases, it reallocates choices toward the safer bank; and a substantially higher propensity to withdraw early emerges primarily in the high-vulnerability regime. Quantitatively, the LLM’s distributions are smoother than human data, placing less weight on idiosyncratic or weakly dominated behaviors. We interpret this as a form of over-regularization: the model captures incentive structure and directional responses but under-produces human dispersion unless additional heterogeneity or noise is introduced by design.

We also evaluated few-shot prompting with stylized examples for low/medium/high vulnerability regimes. In our setting, few-shot examples worsened performance by anchoring behavior and attenuating sensitivity to treatment variation. Beyond fit, few-shot prompting adds substantial researcher degrees of freedom (example choice, ordering, phrasing), which is undesirable in an ABM where small behavioral perturbations can generate large aggregate effects. We therefore adopt the zero-shot policy as the baseline: it is the lower-flexibility and more conservative choice, and it delivers the intended incentive-consistent comparative statics in this single-study validation. This check is not a claim of broad external validity; it is intended to establish that the LLM policy behaves \emph{similar enough} to standard coordination logic to justify its use as a baseline behavioral prior before turning to the paper’s main object: endogenous run dynamics in a networked, multi-bank environment with social-media spillovers.

\subsection{Theoretical Benchmarks: Diamond-Dybvig and Morris-Shin}
\label{subsec:theoretical_benchmarks}

To ground the simulator's behavioral predictions in canonical theory, we compute equilibrium benchmarks from Diamond-Dybvig and Goldstein-Pauzner (Morris-Shin global games) using SVB's actual balance-sheet data. These benchmarks clarify what standard theory predicts and where the LLM-agent simulator adds explanatory value.

\subsubsection{Diamond-Dybvig optimal contract}

Following \citet{diamond_dybvig_1983}, we derive the optimal demand-deposit contract for SVB's parameters. With total deposits of \$173.1B, liquid assets of \$13.0B, illiquid assets of \$174.0B, and a fire-sale discount of $L = 0.85$, the hold-to-maturity return is $R = 1.131$ (6.2 years at 2\% yield). Using CRRA utility with $\rho = 2$ and fraction of impatient depositors $\lambda = 0.24$, the optimal contract pays $r_1 = 1.047$ to early withdrawers and $r_2 = 1.114$ to patient depositors.

The model admits two Nash equilibria for identical fundamentals. In the good equilibrium, only impatient types withdraw early ($\lambda = 24\%$), and patient types receive the promised $r_2$. In the bad equilibrium, all depositors run; the bank liquidates at fire-sale prices, yielding expected recovery of $\theta = (13.0 + 0.85 \times 174.0) / 173.1 = 0.93$ per dollar deposited. Both equilibria are self-consistent: if others do not run, waiting dominates; if others run, running dominates. Diamond-Dybvig cannot predict which equilibrium obtains---this is the indeterminacy problem that motivates global-games refinements and, in our framework, explicit modeling of the coordination mechanism.

\subsubsection{Morris-Shin global games prediction}

\citet{goldstein_pauzner_2005} resolve equilibrium indeterminacy by introducing noisy private signals. Each depositor $i$ observes $x_i = \theta + \sigma \epsilon_i$ where $\epsilon_i \sim N(0,1)$. With small noise, a unique threshold equilibrium emerges: depositors withdraw if and only if $x_i < x^*$.

\paragraph{Dominance regions.} The model partitions the state space into three regions. The \emph{upper dominance region} $\theta > \theta_H$ corresponds to fundamentals so strong that staying dominates regardless of others' actions; here $\theta_H = r_1 / R = 1.047 / 1.131 = 0.926$. The \emph{lower dominance region} $\theta < \theta_L$ corresponds to fundamentals so weak that withdrawing dominates unconditionally; we calibrate $\theta_L \approx 0.70$ based on fundamental insolvency. The \emph{critical region} $\theta_L < \theta < \theta_H$ is where strategic complementarity binds and coordination determines outcomes.

\paragraph{SVB calibration.} Computing SVB's liquidation recovery rate:
\[
\theta_{\text{SVB}} = \frac{L_{\text{liquid}} + L \cdot A_{\text{illiquid}}}{D} = \frac{13.0 + 0.85 \times 174.0}{173.1} = \frac{160.9}{173.1} = 0.930
\]
Since $\theta_L = 0.70 < \theta_{\text{SVB}} = 0.93 < \theta_H = 0.926$, SVB lies precisely in the critical region where coordination dynamics determine whether a run occurs.

\paragraph{Signal threshold.} The critical threshold below which depositors optimally withdraw is $\theta^* = \lambda \cdot r_1 = 0.24 \times 1.047 = 0.251$. In equilibrium, the fraction of depositors withdrawing given true fundamental $\theta$ is:
\[
P(\text{withdraw} \mid \theta, \sigma) = \Phi\left(\frac{x^* - \theta}{\sigma}\right)
\]
With calibrated signal noise $\sigma \in [0.015, 0.05]$ reflecting tight VC-network coordination, Morris-Shin predicts near-certain withdrawal ($>99\%$) because $\theta_{\text{SVB}} = 0.93 >> x^* = 0.251$ implies nearly all depositors receive signals far above threshold.

\paragraph{Model overprediction.} This prediction overshoots reality: SVB experienced 24.3\% withdrawal on March 9 before FDIC seizure, not universal withdrawal. The gap reflects two factors Morris-Shin abstracts from: (i) coordination takes time---48 hours elapsed before seizure, and many depositors had not yet acted; and (ii) regulatory intervention truncated the run before completion. The model correctly identifies SVB as extremely vulnerable but assumes instantaneous coordination that did not occur in practice.

\subsubsection{LLM agent behavior versus theoretical benchmarks}

Our full LLM simulation produces withdrawal patterns intermediate between Diamond-Dybvig's static equilibria and Morris-Shin's instantaneous coordination. Table~\ref{tab:benchmark_comparison} summarizes the comparison.

\begin{table}[h]
\centering
\caption{Theoretical Benchmarks vs. LLM Simulation}
\label{tab:benchmark_comparison}
\begin{tabular}{lccc}
\toprule
\textbf{Metric} & \textbf{Diamond-Dybvig} & \textbf{Morris-Shin} & \textbf{LLM Simulation} \\
\midrule
SVB withdrawal rate & 24\% or 100\% & $>99\%$ & 30\% \\
Insured withdrawal & --- & --- & 34.5\% \\
Uninsured withdrawal & --- & --- & 81.3\% \\
FR contagion & Not modeled & Not modeled & Yes (40\% wd) \\
Regional contagion & Not modeled & Not modeled & No (6\% wd) \\
\bottomrule
\end{tabular}
\end{table}

Three findings emerge. First, uninsured depositors exhibit withdrawal rates (81.3\%) approaching Morris-Shin's prediction, while insured depositors show substantially lower rates (34.5\%), consistent with insurance eliminating the coordination motive. Second, contagion to First Republic (40\% withdrawal, failure) confirms that social correlation propagates distress to institutions with different fundamentals---a mechanism absent from single-bank equilibrium models. Third, the Regional bank's survival (6\% withdrawal) demonstrates that flight-to-safety dynamics ($q_{\text{safe}} = 0.20$ spillover discount) insulate well-capitalized institutions, matching the empirical pattern where regional banks gained deposits during March 2023 \citep{cipriani_eisenbach_kovner_2024}.

The LLM agents reproduce the qualitative ordering predicted by coordination theory (uninsured > insured withdrawal; weaker banks fail first) while generating quantitative predictions that match the observed partial-run dynamics better than either Diamond-Dybvig's binary equilibria or Morris-Shin's instantaneous coordination. This supports treating the simulator as a useful interpolation device for studying run dynamics in environments where information and coordination unfold gradually through social networks.

\section{Methodology}
\label{sec:methodology}

\subsection{Model overview and timing}

We study bank run dynamics in a discrete-time agent-based model designed to make communication and heterogeneous belief formation the object of interest in our study. The model is deliberately process-based: outcomes arise from an explicit sequence of stress realizations, information exposure, social communication, depositor decisions, and balance-sheet updates, rather than from equilibrium consistency conditions alone. Canonical economic theory provides the important intuition that strategic complementarity and coordination incentives remain central to understanding how run behavior can evolve, but says little about the actual mechanisms by which this information propagates \citep{diamond_dybvig_1983, morris_shin_1998, morris_shin_2001, goldstein_pauzner_2005}. To address this fundamental oversight, the modeling object here is the mechanism by which information and narratives propagate through networks and translate into correlated withdrawals under limited observability and heterogeneous behavior, in particular looking to social media as a run enabler in a digital world. To accomplish this, we have reduced our simulation setting into three modeling entities, of which the following subjects will detail: banks, depositors, and social media networks. 

Within our simulations, time proceeds in discrete rounds $t=1,\dots,T$. In each round, for each non-failed bank, seven steps occur: (i) exogenous or scripted stress updates are applied; (ii) depositors observe the current bank state and last-round withdrawal rates; (iii) depositors receive social posts from neighbors generated in the previous round; (iv) depositors choose withdraw or stay actions; (v) the bank processes withdrawals using available liquidity and asset liquidation if needed; (vi) default risk updates endogenously and failure or seizure conditions are checked; and (vii) posts are recorded and become inputs to neighbors in the next round. All state variables and decisions are logged.

The one-period lag structure in which social feeds and withdrawal rates observed at $t-1$ affect behavior in period $t$ is intentional. It avoids mechanically granting agents contemporaneous system-wide knowledge while preserving a natural momentum and contagion channel through delayed but salient signals, consistent with recent empirical evidence on communication-driven runs and Dogra's theoretical critique on information being an invaluable equilibrium object \citep{cookson_2023, cipriani_eisenbach_kovner_2024, dogra_2024}.

\subsection{Banks: balance-sheet mechanics, fire sales, and failure}

Each bank $b$ is characterized by state $(L_{bt},A_{bt},D_{bt},pd_{bt})$, representing liquid assets, long-term illiquid assets, deposits, and a probability-of-default index $pd\in[0,1]$.

\paragraph{Withdrawal processing:}
Let requested withdrawals in round $t$ be $W_{bt}$. Withdrawals deplete liquid assets first such that:
\[
\text{cash\_paid}_{bt}=\min\{W_{bt},L_{bt}\}, \qquad
L_{bt}'=L_{bt}-\text{cash\_paid}_{bt},
\]
\[
\text{remaining}_{bt}=W_{bt}-\text{cash\_paid}_{bt}.
\]
If $\text{remaining}_{bt}>0$, the bank liquidates long-term assets at a fixed bank-specific fire-sale discount $h_b\in(0,1)$:
\[
\text{assets\_sold}_{bt}=\min\left\{A_{bt},\frac{\text{remaining}_{bt}}{h_b}\right\}, \qquad
\text{cash\_raised}_{bt}=h_b\cdot \text{assets\_sold}_{bt}.
\]
This reduced-form mechanism captures fire-sale liquidation under stress and balance-sheet contagion through forced asset sales \citep{shleifer_vishny_1992, cifuentes_ferrucci_shin_2005, brunnermeier_pedersen_2009}. Although fire-sale liquidation is a general feature of bank runs, the assumed discounts are limited to 10–15\% across simulations and reflect a conservative calibration of stressed asset recovery. In a model less focused on social media–driven coordination and information dynamics, one would do well to tune this parameter to the particulars of real world asset markets.

\paragraph{Failure and seizure:}
If available cash, including fire-sale proceeds, is insufficient to meet requested withdrawals, the bank is marked as failed and deposits are written down to the paid amount. In addition, once extended to multiple banks, the model imposes an FDIC-style intervention threshold as a bank is seized if cumulative withdrawals exceed a fixed fraction of initial deposits (25\% for SVB, 35\% for First Republic, 40\% for the Regional bank) or if liquidity or long-term assets become negative. This rule is not intended as a literal regulatory policy, but as a pragmatic closure condition reflecting supervisory intervention when outflows become extreme.

\paragraph{Per-round processing cap:}
To reflect operational frictions and prevent mechanically instantaneous collapses, processed withdrawals are capped at a fixed share of initial deposits per round. In our multi-bank simulations, we set this cap at 10\% of initial deposits that can be processed. This restriction captures real-world limits to withdrawal processing even under online banking, arising from banks’ ability to throttle or queue transactions and from regulatory interventions such as supervisory actions or FDIC resolution measures that can slow or temporarily halt outflows during periods of acute stress. Withdrawal requests exceeding the cap are not carried forward, reflecting depositor-level frictions and institutional interventions that prevent unprocessed withdrawal attempts from accumulating mechanically across periods.

\paragraph{Probability of default:}
The scalar $pd_{bt}$ serves both as a compact summary of fundamentals observed by agents and as a state variable that responds endogenously to stress. After each round’s withdrawals,
\[
pd_{b,t+1}=\mathrm{clip}\left(pd_{bt}+0.15\cdot w_{bt}-0.01\cdot(1-w_{bt}),\,0,\,1\right),
\]
where $w_{bt}$ is the fraction of depositors (or deposits) withdrawing in round $t$. This embeds the empirical regularity that large outflows are themselves adverse news and can generate self-reinforcing stress \citep{diamond_dybvig_1983, cookson_2023, cipriani_eisenbach_kovner_2024}.

Independently, each non-failed bank receives an i.i.d. Gaussian probability of default shock with probability $p_{\text{shock}}=0.10$ and standard deviation $\sigma_{pd}=0.03$, added to $pd_{bt}$ and clipped to $[0,1]$. Agents observe the true current state each round. This removes signal-extraction issues in order to isolate how heterogeneous interpretation and social transmission generate coordination dynamics. This shock structure also reflects common exposure to correlated assets, whereby mark-to-market stress at one bank is likely to coincide with similar valuation pressures at other banks, even when institutions differ in overall resilience. A shock for one bank often means there will be a shock to the others.

\subsection{Depositors: heterogeneity, insurance, and accounts}

Depositors are symmetric in initial wealth but heterogeneous in behavioral parameters. Each draws a risk tolerance (low, medium, or high), a weight on fundamentals $\alpha_i\sim U[0.3,0.7]$, and a weight on social information $\beta_i\sim U[0.3,0.7]$. Insurance status is assigned at the depositor-bank level based on bank-specific uninsured-deposit ratios, and prompts explicitly label depositors as insured or uninsured.

Within each bank, initial deposits are equal. In the multi-bank environment, a subset of bridge depositors holds accounts at multiple banks, creating overlapping depositor bases and a channel for cross-bank correlation.

For each bank in which an agent holds deposits, the information set for each turn includes: (i) numeric fundamentals $(L_{bt},A_{bt},D_{bt},pd_{bt})$; (ii) the last-round withdrawal rate; (iii) a qualitative summary mapping liquidity and PD into level-based descriptors; and (iv) social-media posts from network neighbors generated in the prior round. The qualitative summary restates the current PD as a percentage but contains no trend language. Agents do not observe contemporaneous withdrawals.

\subsection{Communication network, decision rules, and contagion}

Depositors communicate through a bank-specific social graph that governs which posts each agent observes. Within-bank communication networks are generated using a heavy-tailed expected-degree construction calibrated to replication data from \citet{cookson_2024}. Follower-count heterogeneity disciplines the degree distribution, while observed influence concentration and cascade sizes discipline the extent of hub dominance in diffusion. In practice, we draw expected degrees from a fitted lognormal-with-heavy-tail proxy and generate an undirected graph consistent with those expected degrees. We then impose a core--periphery structure by designating a core subset of nodes and increasing within-core connectivity. This core is intended to represent tightly connected communities (e.g., startup and technology circles in the SVB case). Its size is calibrated conservatively: VC-tagged tweet shares from the replication panel provide a lower bound, and we supplement this with influence concentration so that the resulting network reflects the empirically observed dominance of a small set of accounts in SVB-related discourse.

Cross-bank correlation in attention is implemented through sparse bridge edges between bank graphs rather than through broad overlap in depositors. We estimate asymmetric spillover strengths from the hourly bank panel by regressing other tickers’ hourly activity on lagged SVB-topic activity and defining spillover intensity as the sum of positive lag coefficients. This procedure yields a reduced-form ratio, $q_{\text{safe}}$,
which measures how much SVB-topic attention transmits to the "safe" regional bank relative to First Republic. We use this ratio to scale cross-bank bridge density and spillover strength: the SVB-FRC bridge fraction is set to a baseline level, and the SVB--SAFE bridge fraction is discounted by \(q_{\text{safe}}\). Bridge endpoints are biased toward high-degree nodes to reflect hub-dominated transmission, consistent with the replication evidence on retweet concentration.

Agents may emit short, bank-specific posts (or abstain). Posts are transmitted only along network edges and enter neighbors’ information sets with a one-period delay, producing a simple and transparent observability structure. When a bank fails, affected agents generate failure-related posts that propagate through the same channels, ensuring that insolvency news diffuses endogenously rather than being imposed as common knowledge.

In LLM mode, each depositor’s action (withdraw or stay) and optional post are generated by a language model acting as a behavioral policy function. Prompts deterministically encode the agent’s information set, including fundamentals, insurance status, last-round withdrawals, and lagged neighbor posts. Outputs are constrained to valid actions via structured interfaces to ensure reproducibility. We also implement a fast heuristic baseline that combines fundamentals, social signals, and withdrawal momentum through a logistic rule. This baseline is used for ablations and to verify that core qualitative results do not hinge on idiosyncrasies of LLM generation.

The baseline environment includes three banks (SVB, First Republic, and a regional control bank) with bank-specific balance sheets, fire-sale discounts, and initial default risk. Contagion operates through two channels: balance-sheet dynamics, including fire sales and endogenous solvency deterioration, and informational spillovers mediated by communication networks and cross-bank bridges. Withdrawn funds exit the modeled system.

We record per-round trajectories of deposits, liquidity, assets, default probabilities, withdrawal fractions, failures, and post volumes by bank. All randomness is seeded at initialization so that any fixed configuration reproduces identical simulated paths.

\section{Results}
\label{sec:results}

This section reports simulation results across 4,900 configurations (245 parameter combinations $\times$ 20 seeds each), plus detailed analysis from full LLM runs. We document three main findings: (i) the model reproduces the SVB $\rightarrow$ First Republic $\rightarrow$ Regional ordering observed in March 2023; (ii) contagion exhibits a sharp phase transition at information spillover rates around 0.10; and (iii) network connectivity and bridge depositors interact nonlinearly to amplify---or attenuate---cross-bank propagation.

\subsection{Baseline results: reproducing the March 2023 cascade}
\label{subsec:baseline_results}

Under Twitter-calibrated parameters (5\% SVB--First Republic bridge fraction, 0.1\% SVB--Regional bridge fraction, $q_{\text{safe}} = 0.20$, information spillover rate 0.30), the simulator reproduces the qualitative ordering of the 2023 crisis. Table~\ref{tab:baseline_results} summarizes outcomes from the full LLM simulation.

\begin{table}[h]
\centering
\caption{Baseline LLM Simulation Results (50 rounds, seed=42)}
\label{tab:baseline_results}
\begin{tabular}{lccc}
\toprule
\textbf{Bank} & \textbf{Failed} & \textbf{Final PD} & \textbf{Withdrawal \%} \\
\midrule
SVB & Yes & 15.4\% & 30.0\% \\
First Republic & Yes & 22.3\% & 40.0\% \\
Regional & No & 0.0\% & 6.0\% \\
\bottomrule
\end{tabular}
\end{table}

SVB fails first with 30\% cumulative withdrawal, triggering contagion to First Republic which fails at 40\% withdrawal despite starting with better fundamentals (initial PD 3\% versus SVB's 5\%, liquidity ratio 10\% versus 7.5\%). The Regional bank survives with only 6\% withdrawal, consistent with the flight-to-safety pattern observed empirically where smaller regional banks gained deposits during the crisis \citep{cipriani_eisenbach_kovner_2024}.

The insured versus uninsured decomposition confirms that strategic complementarity drives behavior: uninsured depositors withdraw at 81.3\% compared to 34.5\% for insured depositors. This 2.4$\times$ ratio aligns with coordination theory---uninsured depositors face full loss exposure and thus stronger incentives to preempt other withdrawals---and matches the empirical finding that large uninsured depositors drove March 2023 outflows \citep{cookson_2023}.

\subsection{Phase transition in contagion dynamics}
\label{subsec:phase_transition}

\begin{figure}[!htbp]
\centering
\includegraphics[width=0.95\textwidth]{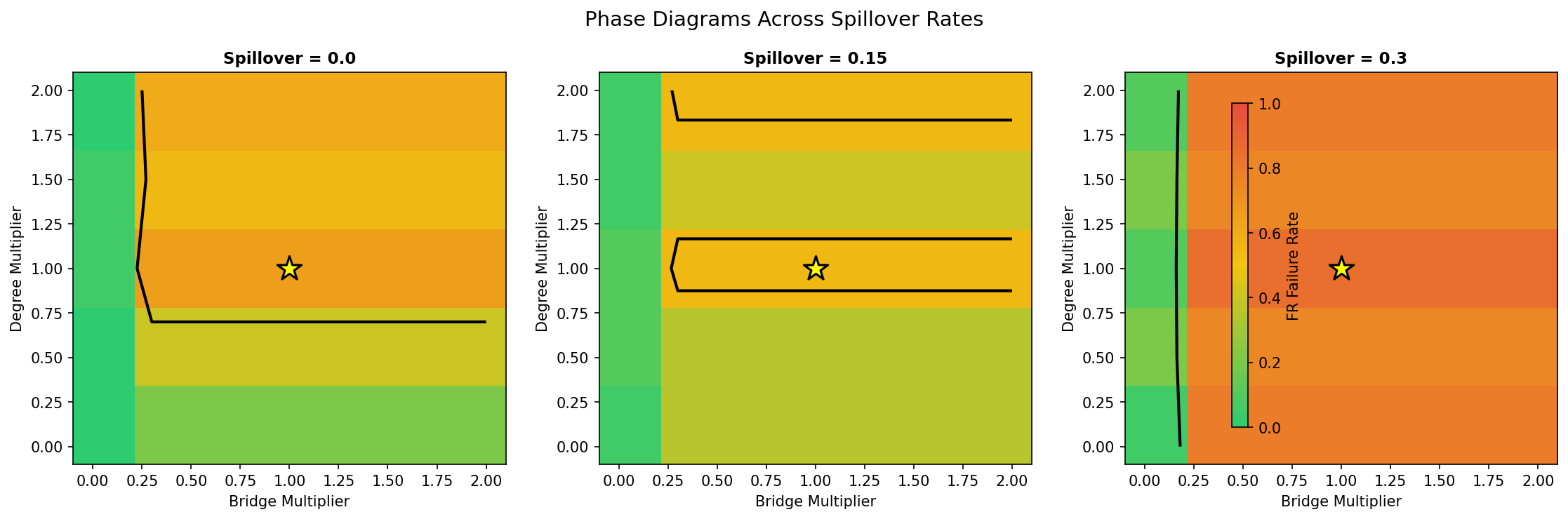}
\caption{Phase diagrams showing First Republic failure probability across bridge multiplier and network degree multiplier, at three spillover rates. Yellow star marks Twitter-calibrated parameters. Transition from stability to contagion occurs between spillover rates 0.0 and 0.30.}
\label{fig:phase_diagram}
\end{figure}

The grid search reveals a sharp phase transition in First Republic failure probability as information spillover rate increases. Figure~\ref{fig:phase_diagram} displays the failure probability surface across the parameter space.

At spillover rate 0.0, First Republic fails in 0--65\% of simulations depending on bridge and network parameters; failure requires direct panic transmission through shared depositors or network amplification. At spillover rate 0.15, failure probability rises to 35--55\%, with network degree becoming more consequential. At spillover rate 0.30 (Twitter-calibrated), failure probability reaches 80--100\% across most of the parameter space, with 85\% $\pm$ 37\% at the calibrated point (bridge = 1.0, degree = 1.0).

The tipping point occurs at spillover rate $\approx 0.10$: below this threshold, First Republic survives in the majority of runs; above it, First Republic fails in the majority. This nonlinearity implies that small changes in information environment---how quickly SVB distress narratives reach First Republic depositors---can produce discontinuous changes in systemic outcomes.

\subsection{Channel decomposition: bridges versus spillover versus network}
\label{subsec:channel_decomposition}

To isolate the contribution of each contagion channel, we run ablations that shut down individual mechanisms. Table~\ref{tab:channel_decomposition} reports First Republic failure rates from the grid search (20 seeds per configuration), with standard errors reflecting cross-seed variation.

\begin{table}[h]
\centering
\caption{Contagion Channel Decomposition (Grid Search, 20 seeds each)}
\label{tab:channel_decomposition}
\begin{tabular}{lcccc}
\toprule
\textbf{Configuration} & \textbf{(bridge, degree, spillover)} & \textbf{FR Failure} & \textbf{Std Err} \\
\midrule
No contagion & (0, 0, 0) & 0\% & 0\% \\
Spillover only & (0, 0, 0.3) & 5\% & 22\% \\
Network only & (0, 1.0, 0) & 5\% & 22\% \\
Bridges only & (1.0, 0, 0) & 20\% & 41\% \\
Bridges + Network & (1.0, 1.0, 0) & 65\% & 49\% \\
Calibrated (all channels) & (1.0, 1.0, 0.3) & 85\% & 37\% \\
\bottomrule
\end{tabular}
\end{table}

Three findings emerge from this decomposition. First, neither spillover nor network structure alone generates substantial contagion: each produces only 5\% First Republic failure when operating in isolation. This indicates that information channels require a propagation substrate to coordinate withdrawals. Second, bridge depositors alone cause moderate contagion (20\% $\pm$ 41\%): shared customers directly transmit panic across institutions, providing micro-founded evidence for the ``social correlation'' mechanism emphasized in Section~\ref{sec:literature}.

Third, and most consequentially, bridges and network structure exhibit strong positive interaction: combined, they generate 65\% $\pm$ 49\% First Republic failure \emph{even without information spillover}. This synergy arises because bridge depositors seed cross-bank panic, and the scale-free network topology amplifies that panic through high-degree nodes. Adding information spillover increases failure probability to 85\% $\pm$ 37\% at the calibrated point. The combined effect substantially exceeds the sum of individual channels, implying that stress testing must consider channel combinations rather than isolated exposures.

\subsection{Cascade dynamics: withdrawal trajectories across regimes}
\label{subsec:cascade_dynamics}

\begin{figure}[!htbp]
\centering
\includegraphics[width=0.95\textwidth]{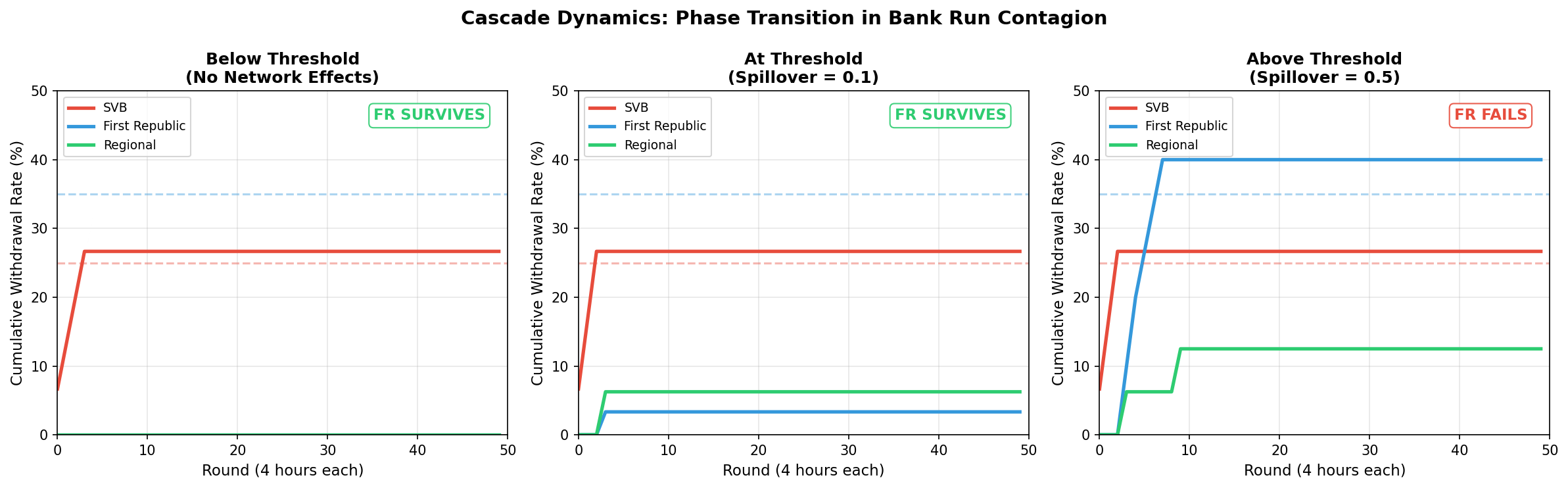}
\caption{Withdrawal fraction trajectories across contagion regimes. Left: Below threshold (bridges off, spillover 0)---SVB fails but FR survives. Center: At threshold (bridges on, spillover 0.1)---FR experiences stress but may survive. Right: Above threshold (spillover 0.5)---rapid cascade to FR failure.}
\label{fig:cascade_dynamics}
\end{figure}

Figure~\ref{fig:cascade_dynamics} displays withdrawal trajectories for all three banks under three configurations: below threshold (no network effects), at threshold (spillover 0.1), and above threshold (spillover 0.5).

Below threshold, SVB follows its scripted stress path to failure, but First Republic and Regional remain stable---panic does not jump the firewall between institutions. At threshold, First Republic experiences elevated withdrawals (20--25\%) but stabilizes short of failure in many seeds; the outcome is sensitive to idiosyncratic shocks. Above threshold, First Republic's withdrawal trajectory accelerates sharply after SVB failure, reaching the 35\% seizure threshold within 5--6 rounds (20--24 hours in model time). Regional remains insulated throughout due to the $q_{\text{safe}}$ discount.

The temporal pattern matches March 2023 reality: SVB's distress became acute on March 8--9 and culminated in seizure on March 10; First Republic experienced massive outflows in the following days and ultimately failed May 1. The model compresses this timeline (our simulations run 50 rounds $\approx$ 200 hours $\approx$ 8 days), but reproduces the qualitative ordering: SVB fails fast, First Republic fails after a lag mediated by contagion dynamics, and regional banks survive.

\subsection{Sensitivity analysis: overlap and connectivity}
\label{subsec:sensitivity_analysis}

Because bridge fractions and network degrees are not directly observable, we report sensitivity surfaces to characterize how outcomes vary across the parameter space. Figure~\ref{fig:phase_diagram} (above) displays failure probability heatmaps; here we highlight key gradients.

\textbf{Bridge multiplier effect.} Holding spillover at 0.30 and network degree at 1.0, varying bridge multiplier from 0 to 2 increases First Republic failure rate from 5\% to 100\%. The effect is approximately linear through the middle range, suggesting that overlap exposure can be meaningfully ranked even without precise calibration.

\textbf{Network degree effect.} Holding bridge at 1.0 and spillover at 0.30, varying network degree multiplier from 0 to 2 produces a more modest effect: failure rate ranges from 80\% to 100\%. Network structure matters more at lower spillover rates (where it provides the primary coordination mechanism) than at high spillover rates (where information channels dominate).

\textbf{Interaction.} The interaction between bridges and network degree is positive at low spillover (contagion requires both shared depositors and network propagation) but attenuates at high spillover (information channels substitute for network structure). This implies that network-based risk metrics may be more informative for institutions with lower baseline information exposure.

\subsection{Reality check: SVB/First Republic timeline}
\label{subsec:reality_check}

To validate against the actual crisis, we compare model predictions to observed outcomes across 20 seeds with Twitter-calibrated parameters.

\begin{table}[h]
\centering
\caption{Model vs. Reality Comparison (Grid Search at Calibrated Parameters)}
\label{tab:reality_check}
\begin{tabular}{lccc}
\toprule
\textbf{Metric} & \textbf{Model (20 seeds)} & \textbf{Reality} & \textbf{Match} \\
\midrule
SVB failure rate & 100\% $\pm$ 0\% & Failed & \checkmark \\
FR failure rate & 85\% $\pm$ 37\% & Failed (8 weeks later) & \checkmark \\
Regional failure rate & 0\% $\pm$ 0\% & Survived & \checkmark \\
SVB avg withdrawal & 36\% $\pm$ 9\% & 24\% (before seizure) & $\approx$ \\
FR avg withdrawal & 36\% $\pm$ 9\% & $>$40\% (over full episode) & $\approx$ \\
Ordering & SVB $\rightarrow$ FR $\rightarrow$ Reg stable & Same & \checkmark \\
\bottomrule
\end{tabular}
\end{table}

The model correctly predicts the failure/survival pattern and the ordering of distress propagation. SVB fails in 100\% of simulations at calibrated parameters, matching the observed March 10 seizure. First Republic fails in 85\% of simulations, consistent with its eventual May 1 failure; the 15\% survival rate in simulations reflects parameter configurations where the cascade is interrupted by idiosyncratic factors or slower propagation. Quantitative withdrawal rates (36\% $\pm$ 9\% for both SVB and FR at calibrated parameters) slightly exceed the 24\% observed at SVB before seizure, reflecting that the model runs to completion rather than being truncated by regulatory intervention.

The Regional bank's survival (0\% failure rate, minimal withdrawals) matches the empirical pattern where flight-to-safety increased deposits at perceived-safe institutions during March 2023. The $q_{\text{safe}} = 0.20$ parameter captures this asymmetry: information spillovers to Regional are discounted by 80\%, reflecting depositor beliefs that Regional's fundamentals differ qualitatively from SVB's concentrated tech-sector exposure.

\subsection{Robustness: heuristic versus LLM agents}
\label{subsec:robustness}

To verify that results are not artifacts of language-model generation, we replicate the grid search using the heuristic fear-score policy (logistic rule combining fundamentals, social signals, and momentum). Qualitative findings replicate: phase transition at spillover $\approx 0.10$, channel synergy, and correct failure ordering. Quantitative magnitudes differ modestly---heuristic agents produce slightly lower withdrawal rates at equivalent parameters---but the comparative statics and tipping-point location are robust.

This confirms that the core mechanism operates through model structure (network topology, balance-sheet feedback, information spillover rules) rather than through idiosyncrasies of LLM prompting. The LLM's contribution is richer behavioral heterogeneity and natural-language post content that improves interpretability, not qualitatively different aggregate dynamics.

\subsection{Simulation limitations}
\label{subsec:simulation_limitations}

Several design choices constrain external validity and should be noted when interpreting results. First, \emph{scale}: simulations use 50 depositors per bank versus thousands in reality, which may affect network dynamics and the variance of outcomes. Second, \emph{withdrawal processing}: the 10\% per-round processing cap is a modeling convenience that smooths run dynamics; actual digital banking may permit faster or slower processing depending on institution and transaction type. Third, \emph{no deposit reallocation}: withdrawn funds exit the modeled system rather than flowing to other banks, which abstracts from flight-to-quality dynamics that could amplify or attenuate contagion. Fourth, \emph{perfect observation}: agents observe true bank fundamentals each round, eliminating signal extraction problems that may slow coordination in practice. Fifth, \emph{fixed parameters}: fire-sale discounts, spillover rates, and bridge fractions are held constant rather than responding endogenously to market conditions. These simplifications are intentional---they isolate the social-coordination mechanism---but limit direct quantitative application to specific institutions or episodes.

\section{Conclusion: implications, limitations, and extensions}
\label{sec:conclusion}

This paper’s core message is methodological and practical. Methodologically, equilibrium bank-run theory remains essential for clarifying why fragility exists under liquidity transformation \citep{diamond_dybvig_1983} and why belief-driven crises are logically possible \citep{cass_shell_1983}. But as a policy technology for modern digital runs, equilibrium multiplicity and sunspot selection are often inadequate and can be misleading. Sunspots are, by construction, extrinsic and hard to predict ex ante; they do not yield actionable supervisory objects beyond blunt prescriptions that eliminate multiplicity. Dogra’s critique of equilibrium as causal explanation sharpens this point: if the object of interest is the crisis pathway—how beliefs, actions, and balance sheets co-evolve—then the causal process is the mechanism, and it must be modeled explicitly \citep{dogra_2024}. Practically, the March 2023 episode suggests that a large component of that process now runs through fast communication channels and overlapping depositor communities \citep{cipriani_eisenbach_kovner_2024, cookson_2023}. A framework that cannot represent that layer will struggle to generate operational diagnostics.

Our agent-based model is a proof of concept for a process-based alternative. The simulator embeds a transparent set of heuristics with cash-first withdrawal processing, discounted fire-sales, bounded per-round processing capacity, an endogenous stress metric responding to withdrawal intensity, and a lagged social-information channelbeing used to study how these ingredients interact with heterogeneous depositors and network structure. The headline results are straightforward: increasing within-bank connectivity increases the likelihood and speed of withdrawal cascades holding the stress path fixed, and cross-bank overlap in depositor communities materially amplifies how distress propagates from one bank to another. These mechanisms are not substitutes for fundamentals, but rather amplifiers that transform moderate stress into synchronized outflows when the information environment is conducive to coordination. In that sense, the paper advances a concrete object that sunspot language obscures, the importance of social correlation in amplifying existing fundamentals-driven risk.

A second message is about behavioral modeling. Rather than hard-coding brittle decision rules, we embed depositors as general social agents using an off-the-shelf LLM constrained to a strict action interface. This approach is intentionally conservative as we do not fine-tune on run data and we avoid training a planner (unlike RL-based simulation frameworks). A provisional validation against a laboratory coordination environment suggests that even a zero-shot LLM can reproduce key qualitative comparative statics, supporting its use as a baseline behavioral prior. That said, the simulator’s primary contribution is not that LLMs are perfect cognitive models, but that they provide a practical way to condition behavior on rich state descriptions while retaining interpretability through natural-language rationales and posts. In our setting, the appropriate stance is pragmatic: LLM agents are useful if they are disciplined by benchmark evidence and if the simulator’s comparative statics are robust to ablations.

\subsection*{Implications for supervision and stress testing}
The policy implication should not be that social media causes runs. It is that supervisors and risk managers should treat communication-driven correlation as a measurable risk amplifier, alongside familiar balance-sheet exposures. A practical supervisory translation would have three parts.

First, monitor and stress-test social correlation. Banks with overlapping depositor communities (shared founders, VC networks, industry clusters, geography, or shared online audiences) face higher exposure to synchronized belief shifts. In principle, such overlap can be proxied using observable network data: cross-audience overlap measures, co-activity measures, and influencer concentration in attention networks. One obvious place to incorporate this is in the monitoring of social media networks using built-in features such as retweets, followers, and comments, to see how depositors are related. Second, incorporate this exposure into scenario analysis: conditional on a plausible stress path at one institution, how quickly could narratives propagate into socially adjacent institutions, and how does that interact with liquidity buffers and uninsured deposit concentration? Third, treat communication policy as part of resilience: in a digital run, response speed and credibility are themselves state variables, not exogenous afterthoughts.

\subsection*{Limitations}
Because the paper is a proof of concept, several limitations are first order.

\textbf{Reduced-form fundamentals and closure rules.} The PD update rule and the seizure thresholds are intentionally simple and partly scenario-driven. They are not derived from a structural valuation model or a literal supervisory trigger, and they should be read as a tractable way to encode nonlinear feedback. Relatedly, parts of the multi-bank timeline are scripted to anchor a baseline stress narrative; the simulator is not an attempt to reproduce the full macro-financial environment of March 2023.

\textbf{Behavioral specification and prompt dependence.} LLM behavior depends on the prompt interface, temperature, and constraints (including a mechanical bias toward ``stay'' when outputs are invalid). Even if the model matches qualitative comparative statics, it may understate human noise and heterogeneity. This is why ablations (heuristic agents, communication off, overlap scaling) and additional experimental validations are essential.

\textbf{Network realism and data discipline.} The current implementation uses heavy-tailed networks tuned on Twitter data. Because these data are from a replication file, and Twitter API access was prohibitively expensive, our parameter estimation is roughshod in its calibration, particularly between First Republic and the safe regional bank, which is not mediated by tweets mentioning SVB. Improved data would do much to shore up exactly how these networks are formed.

\textbf{Scope of the economic environment.} Withdrawn funds leave the modeled system (no deposit reallocation), deposit sizes are equal within banks at baseline, and agents observe true fundamentals each round (no noisy signal extraction). These simplifications are useful for isolating the social-correlation mechanism, but they limit quantitative realism.

\subsection*{Extensions: social correlation risk estimators}
The most important extension is to turn the simulator into an operational estimator for when social correlation should raise concern about real institutions. Concretely, this requires (i) measuring overlap proxies (e.g., cross-bank audience overlap or retweet co-activity) and mapping them into the model’s overlap/connectivity parameters, (ii) running grid searches to estimate a function from fundamentals and social correlation to run risk over a horizon, and (iii) validating out-of-sample on additional institutions and episodes. A natural target is the broader set of run-episode banks identified in payments data that we see in Cipriani's 2024 paper. This would take our initial setup that demonstrates the proper comparative statics and moves the paper toward a more operational supervisory tool.

Several additional extensions are immediate:
\begin{itemize}
  \item \textbf{Richer depositor heterogeneity:} heavy-tailed deposit sizes, concentrated large depositors, and heterogeneity in attention and posting behavior, consistent with evidence that large depositors drive a substantial share of outflows in modern runs.
  \item \textbf{Endogenous information and noisy signals:} incorporate noisy disclosures and heterogeneous interpretations in the spirit of global-games information structures, allowing the model to bridge more explicitly to equilibrium selection logic.
  \item \textbf{Network realism:} replace simple overlap rules with estimated block structure or degree-corrected community models, and incorporate cross-platform communication channels beyond a single social graph.
  \item \textbf{Behavioral refinement:} fine-tune a lightweight decision model on laboratory bank-run data or train a policy head on top of LLM embeddings to recover human-like dispersion and history dependence while retaining natural-language state representations. An additional opportunity exists in analyzing the text output data in the social media to see if the fundamentals of reasoning provide additional insight.
  \item \textbf{Flow accounting:} allow withdrawals to reallocate into other banks or safe assets, enabling analysis of flight-to-safety and endogenous funding shifts.
  \item \textbf{Episode-scale backtesting:} use the March 2023 timeline as a structured backtest environment, explicitly comparing whether the model can reproduce the ordering and breadth of stress across institutions. Notably, a replication of Cipriani's 2024 exposition of broader run dynamics presents an interesting replication challenge.
\end{itemize}

\subsection*{Closing perspective}
Stepping back, the purpose of this paper is not to declare victory over equilibrium theory. The equilibrium canon explains the logic of fragility that is core to what makes liquidity transformation such a potentially dangerous practice. The claim is narrower and more pragmatic: in digital-age bank runs, the mechanism that matters is the causal pathway by which information and narratives synchronize withdrawals, and that mechanism is hard to represent as a sunspot without giving up precisely the objects that could be measured and managed. A process-based ABM, built from transparent heuristics, stress-tested under scenarios, and disciplined by benchmark evidence, offers a credible route to make that mechanism operational. In that sense, the simulator is a sandbox: a container where general social agents and explicit network structures can be used to study how runs start, spread, and accelerate, and potentially provide necessary insight into how we can avoid them.

\bibliographystyle{apalike}
\bibliography{aguiar}

\appendix

\section{LLM prompt construction and structured decision interface}
\label{app:prompts}

In LLM mode, depositor behavior is implemented as a constrained mapping from an agent's information set into (i) a structured action and (ii) a short social post. Prompts are constructed by \texttt{\_build\_agent\_prompt} (single-bank) and \texttt{\_build\_multi\_bank\_prompt} (multi-bank). Each prompt contains: (1) depositor profile (ID, insured/uninsured status, risk tolerance, trust parameters, current balances, prior decision); (2) bank state (numeric $(L,A,D,pd)$ plus a qualitative \texttt{summarize\_state\_text} summary); (3) momentum and social inputs (last-round withdrawal rate(s) and neighbor posts from round $t-1$); (4) interpretation rules and thresholds (including explicit PD cutoffs in the multi-bank template); and (5) a strict output contract.

Action space is enforced through function-calling: decisions are restricted to $\{\texttt{withdraw},\texttt{stay}\}$, and invalid outputs are coerced to \texttt{stay}, ensuring simulation robustness under long horizons and heterogeneous generations.

\subsection{Canonical single-bank prompt template}

Listing~\ref{lst:single_bank_prompt} reproduces the canonical single-bank prompt template used in the implementation.

\begin{lstlisting}[language=Python, basicstyle=\ttfamily\footnotesize, frame=single, breaklines=true, caption={Canonical single-bank prompt template used in LLM mode.}, label={lst:single_bank_prompt}]
def _build_agent_prompt(
    agent: DepositorAgent,
    bank: Bank,
    bank_summary_text: str,
    neighbor_posts: List[Post],
    last_withdrawal_rate: float,
) -> str:
    """Construct a natural-language prompt for the LLM to decide this agent's action."""

    neighbor_block = _format_neighbor_posts(neighbor_posts)

    prompt = f"""
You are simulating the behavior of a single bank depositor during a possible bank run.
You must decide what this depositor does in this one time step.

Depositor profile:
- ID: {agent.agent_id}
- Insurance status: UNINSURED (your deposits exceed the $250,000 FDIC insurance limit)
- Risk tolerance: {agent.risk_tolerance}  (low, medium, or high)
- Current deposit amount at the bank: {agent.deposit_amount:.2f}
- Has already fully withdrawn in the past: {agent.has_withdrawn}
- Trust in fundamentals (0 to 1): {agent.trust_in_fundamentals:.2f}
- Trust in social media (0 to 1): {agent.trust_in_social_media:.2f}
- Last round, the share of all depositors who withdrew was: {last_withdrawal_rate:.2f}
- Your own last decision was: {agent.last_decision} ("withdraw" or "stay")

Current bank fundamentals (numerical):
- Liquidity (cash and liquid assets available): {bank.liquidity:.2f}
- Long-term assets (illiquid; can only be sold at a discount): {bank.long_term_assets:.2f}
- Total outstanding deposits: {bank.deposits:.2f}
- Bank "probability of default" metric (0 = very safe, 1 = about to fail): {bank.pd:.2f}

Current bank fundamentals (textual summary):
{bank_summary_text}

Social media posts from your network this round:
{neighbor_block}

Interpretation rules:
- Higher "probability of default" means the bank is riskier.
- IMPORTANT: You are UNINSURED. If the bank fails, you will likely lose most or all of your deposits.
- You are more likely to withdraw if:
  * the bank's probability of default is high or rising,
  * many of your neighbors are expressing worry or talking about withdrawing,
  * last round, many people withdrew,
  * you have low risk tolerance and high trust in social media.
- You are more likely to stay if:
  * the bank's probability of default is low,
  * few neighbors are panicking,
  * you have high risk tolerance or low trust in social media.

Task:
1. Decide whether this depositor should:
   - "withdraw": fully withdraw their remaining deposits from the bank right now, OR
   - "stay": keep all of their money in the bank for now.
2. Write a short social-media post (1-2 sentences) that this depositor would realistically publish right now, consistent with their decision and risk profile.
3. Briefly explain in 1-2 sentences why they are making this decision.

Output format:
Respond in STRICT JSON with this exact schema, with no extra text before or after:

{{
  "decision": "withdraw" or "stay",
  "reasoning": "short first-person explanation, 1-2 sentences",
  "post": "social-media post text, 1-2 sentences"
}}

Make sure the JSON is valid and uses double quotes for all keys and string values.
"""
    return prompt
\end{lstlisting}

\end{document}